\documentclass[11pt,a4paper]{article}
\usepackage{jheppub}
\usepackage{amsmath}
\usepackage{multicol,bbm}
\usepackage{enumerate}
\usepackage{bibentry}
\usepackage{epstopdf}

\usepackage{amssymb}
\usepackage{bm}
\usepackage{multirow}
\usepackage{diagbox}
\usepackage{caption}
\usepackage{tabularx}

\def\be{\begin{equation}}
\def\ee{\end{equation}}
\def\ba{\begin{eqnarray}}
\def\ea{\end{eqnarray}}
\def\nl{\nonumber\\}

\def\a{\alpha}
\def\b{\beta}
\def\c{{\cdot}}
\def\d{\delta}

\def\l{\langle}

\def\b#1{\overline{#1}}

\def\CP1{\mathbb{CP}^1}
\def\SL2C{\mathrm{SL}(2,\mathbb{C})}

\def\Z2{\mathbb{Z}_2}

\def\su2{{SU(2)}}

\def\a{{\alpha}}

\def\[{\left[}
\def\]{\right]}

\def\l{\lambda}
\def\e{\epsilon}

\def\s{\sigma}
\def\a{\alpha}
\def\b{\beta}

\def\({\left(}
\def\){\right)}
\def\[{\left[}
\def\]{\right]}

\def\<{\langle}
\def\>{\rangle}

\def\i2{\frac{i}{2}}

\def\2F1{\,_2{\rm F}_1}

\begin{document}


\title{New Formulas for Amplitudes from Higher-Dimensional Operators}


\author[a,b]{Song He,}
\author[c,a]{Yong Zhang}
\affiliation[a]{CAS Key Laboratory of Theoretical Physics, Institute of Theoretical Physics, Chinese Academy of Sciences, Beijing 100190, China}
\affiliation[b]{School of Physical Sciences, University of Chinese Academy of Sciences, No.19A Yuquan Road, Beijing 100049, China}
\affiliation[c]{Department of Physics, Beijing Normal University, Beijing 100875, China}
\emailAdd{songhe@itp.ac.cn, yongzhang@itp.ac.cn}

\date{\today}

\abstract{In this paper we study tree-level amplitudes from higher-dimensional operators, including $F^3$ operator of gauge theory, and $R^2$, $R^3$ operators of gravity, in the Cachazo-He-Yuan formulation. As a generalization of the reduced Pfaffian in Yang-Mills theory, we find a new, gauge-invariant object that leads to gluon amplitudes with a single insertion of $F^3$, and gravity amplitudes by Kawai-Lewellen-Tye relations. When reduced to four dimensions for given helicities, the new object vanishes for any solution of scattering equations on which the reduced Pfaffian is non-vanishing. This intriguing behavior in four dimensions explains the vanishing of graviton helicity amplitudes produced by the Gauss-Bonnet $R^2$ term, and provides a scattering-equation origin of the decomposition into self-dual and anti-self-dual parts for $F^3$ and $R^3$ amplitudes. }

\maketitle

\section{Introduction and Motivations}

Higher-dimensional operators in gauge theory and gravity are important for various reasons: they are of phenomenological interests as potential corrections to Yang-Mills and Einstein theory; they can appear in effective actions of open and closed strings, and serve as potential counter terms for UV divergences of loop amplitudes. The simplest gauge-invariant, local operator that one can add to Yang-Mills action is the $F^3$ operator,
\be
F^3\equiv \mathrm{Tr}(F_{\mu}^{\;\nu}F_{\nu}^{\;\rho}F_{\rho}^{\;\mu})
=\frac{1}{2}f^{abc}\,F_{\mu}^{a\nu}F_{\nu}^{b\rho}F_{\rho}^{c\mu}\;,
\ee
where $F_{\mu \nu} \equiv F^a_{\mu \nu} T^a$ is the gluon field strength, and $f_{abc}=\mathrm{Tr}([T^a,T^b]T^c)$ the structure constant of gauge group.
This operator arises as the first correction to Yang-Mills Lagrangian $F^2\equiv \mathrm{Tr}(F^{\mu \nu} F_{\mu \nu})$, from the $\alpha'$-expansion of bosonic open string theory~\cite{Polchinski:1998rq}. It is the unique, CP-even dimension-six operator from gauge fields, and it is not supersymmetrizable. The amplitude produced by $F^3$ differs significantly from those produced by higher-dimensional operators in open superstrings. The polarization dependence of the latter is like in Yang-Mills case, {\it e.g.} no contractions of the form $(\epsilon\cdot k)^n$~\cite{Barreiro:2012aw, Barreiro:2013dpa, Boels:2016xhc}, but amplitudes produced by $F^3$ certainly contain such contractions. In this sense, $F^3$ is the first higher-dimensional operator with genuinely new polarization structures in the amplitudes.

$F^3$ operator represents a possible deviation of gluon interactions from those in QCD, which could be produced by new physics~\cite{Simmons:1989zs,Simmons:1990dh,Cho:1993eu}. There have been phenomenological studies on the effect of $F^3$-modified amplitudes~\cite{Simmons:1989zs,Simmons:1990dh,Cho:1993eu,Duff:1991ad,Dreiner:1991xi,Dixon:1993xd}, which were systematically computed using MHV vertex expansion in~\cite{Dixon:2004za} (for BCFW recursions see \cite{Bianchi:2014gla}). In the following we denote the matrix element with $n$ gluons and a single insertion of $F^3$ as $M^{F^3}_n$\footnote{The effective Lagrangian is ${\cal L}= F^2+ \alpha' F^3+ {\cal O}(\alpha'^2)$. We strip off the coupling $g^{n{-}2}$ for pure Yang-Mills amplitude $M^{\rm YM}_n$, and $3 \alpha' g^{n{-}2}$ for $M^{F^3}_n$. The $F^3$ modifications do not change the group theory structure of Yang-Mills action, and in particular the color decomposition of $M^{F^3}_n$ is identical to $M^{\rm YM}_n$.}.

Furthermore, in~\cite{Broedel:2012rc}, it has been argued that $M^{F^3}_n$ satisfies a duality between color and kinematics first proposed for $M^{\rm YM}_n$~\cite{Bern:2008qj}, then double-copy constructions~\cite{Bern:2008qj} give gravity amplitudes from the low-energy effective action of the bosonic closed strings, up to ${\cal O}(\alpha'^2)$:
\ba
S=-\frac{2}{\kappa^2}\int d^4 x\sqrt{g}\,\big[R-2(\partial_{\mu}\phi)^2-\frac{1}{12}H^2+\a'\frac{1}{4}e^{-2\phi}G_2 +\a'^2e^{-4\phi}\big(\frac{1}{48}I_1+\frac{1}{24}G_3\big)+{\cal O}(\a'^3)\Big]\;,\nonumber
\ea
where $G_2$ is the usual Gauss-Bonnet term that contains two powers of Riemann tensor and we will refer it as $R^2$, $I_1$ and $G_3$ contain three powers of Riemann tensors (see~\cite{Broedel:2012rc} for details). If we restrict ourselves to pure gravitons, then at ${\cal O}(\alpha')$ the amplitude is produced by the $R^2$ operator only, but at ${\cal O}(\alpha'^2)$ it receives contribution both from $R^3$ operator as well as two insertions of $R^2$ operators with exchange of a dilation $\phi$. Nevertheless, in the following we will refer to gravity amplitudes from the effective action at ${\cal O}(\alpha')$ and ${\cal O}(\alpha'^2)$ as the $R^2$ and $R^3$ amplitudes, respectively.

Equivalent to the double-copy construction, the corresponding amplitudes can be obtained from those in open strings using field-theory limit of Kawai-Lewellen-Tye (KLT) relations~\cite{Kawai:1985xq}\footnote{Up to ${\cal O}(\alpha'^2)$ only field-theory KLT relations are needed since the stringy corrections start at ${\cal O}(\alpha'^3)$.}. Given that $M^{F^3}_n$ is the ${\cal O}(\alpha')$ correction to $M^{\rm YM}_n$ in open string theory, $M^{R^3}_n$ at ${\cal O}(\alpha'^2)$ comes from double-copy/KLT of two copies of $M^{F^3}_n$, while $M^{R^2}_n$ at ${\cal O}(\alpha')$ can be obtained as the double-copy/KLT of $M^{F^3}_n$ with $M^{\rm YM}_n$.

In four dimensions, it is natural to split the field strength $F$ into self-dual and anti-self-dual parts $F_{\pm}^{\mu\nu}=F^{\mu\nu} \pm \tilde{F}^{\mu\nu}$, and we have amplitudes produced by $F_+^3$ and $F_-^3$ accordingly. The only possible modification to the three-point on-shell gluon amplitudes are the $F^3_+$ amplitude for helicities $(-,-,-)$ and the $F^3_-$ one for $(+,+,+)$:
\be
M^{F_+^3}_3 (-,-,-)=\langle 1\,2\rangle \langle 2\,3\rangle \langle 3\,1\rangle\,,\qquad M^{F_-^3}_3 (+, +, +)=[1\,2][2\,3][3\,1]\,,
\ee
while for any other helicities $F^3$ amplitudes vanish.  $R^3$ amplitudes at ${\cal O}(\alpha'^2)$ are the squaring of $A^{F^3}_3$, and it is important to note that pure graviton amplitudes in four dimensions are from two copies of gauge-theory amplitudes with identical helicities:
\be
M_3^{R^3_+}(-,-,-)=(\langle 1\,2\rangle \langle 2\,3\rangle \langle 3\,1\rangle)^2\,,\qquad M_3^{R^3_-} (+,+,+)=([1\,2][2\,3][3\,1])^2\,.
\ee
On the other hand, $R^2$-modified three-graviton amplitude vanishes for any helicities because $M_3^{R^2}=M^{F^3}_3 \times M_3^{\rm YM}$ and $M_3^{\rm YM}$ is non-vanishing only for $(-,-,+)$ and $(+,+,-)$. In fact, $M_n^{R^2}$ vanishes  for any number of gravitons in four dimensions, because there $R^2$ is a total derivative and cannot produce non-vanishing matrix element. This immediately gives a very interesting relation observed in~\cite{Broedel:2012rc}, namely in four dimensions the KLT of $M^{F_3}_n$ and $M^{\rm YM}_n$ with same helicity configurations must vanish. A similar relation also observed in~\cite{Broedel:2012rc} is that the KLT of $M^{F_+^3}_n$ and $M^{F_-^3}_n$ with same helicities also vanishes:
\be\label{4drel}
M^{F^3}_n (\epsilon^{\pm}) \otimes_{\rm KLT} M^{\rm YM}_n  (\epsilon^{\pm})=0\,,\quad M^{F_-^3}_n  (\epsilon^{\pm} ) \otimes_{\rm KLT} M^{F_+^3}_n  (\epsilon^{\pm})=0\,,
\ee
where $\otimes_{\rm KLT}$ means combining two sets of gauge-theory amplitudes via KLT relations reviewed below, and every pair of gluons must have polarizations with identical helicity, $\epsilon^{\pm}$. For general $n$, these are highly non-trivial relations or $F^3$ amplitudes in four dimensions.

In this paper we study these amplitudes from higher-dimensional operators in the Cachazo-He-Yuan (CHY) formulation~\cite{Cachazo:2013gna,Cachazo:2013hca}. It expresses tree-level S-matrices of massless particles as integrals over the moduli space of punctured Riemann spheres, and naturally incorporates a large variety of theories~\cite{Cachazo:2013iea, Cachazo:2014nsa, Cachazo:2014xea}. As we will review shortly, in the formula for Yang-Mills or Einstein gravity, the most important ingredient is the reduced Pfaffian (or determinant) of a matrix $\Psi_n (\epsilon)$, with manifest gauge/diffeomorphism invariance. The reduced Pfaffian encodes polarization dependence of amplitudes in Yang-Mills, as well as from higher-dimensional operators of open superstrings.

We will present remarkably simple formulas for $M^{F^3}_n$, $M_n^{R^3}$ and $M_n^{R^2}$, which are related to each other through KLT/double-copy constructions. The formulas are all based on one new, gauge-invariant ingredient, ${\cal P}_n$, constructed from the same matrix $\Psi_n (\epsilon)$ with mass dimension higher than the reduced Pfaffian by two. Just as the reduced Pfaffian being the basic object for gluon amplitudes with supersymmetries, ${\cal P}_n$ can be regarded as the basic object for non-supersymmetrizable operators, at least for this lowest dimension.

Furthermore, we study ${\cal P}_n$ in four dimensions, where any CHY formula naturally becomes a sum of contributions from different {\it sectors}~\cite{Cachazo:2013iaa, He:2016vfi}. Given any helicity configuration, it has been known for some time that the reduced Pfaffian is only non-vanishing in one particular sector. This reproduces various twistor string formulas for (super)Yang-Mills and gravity amplitudes~\cite{Witten:2003nn,Roiban:2004yf, Cachazo:2012da, Cachazo:2012kg,Cachazo:2012pz}. As we will see shortly, the reduction of ${\cal P}_n$ to four dimensions is very different: it vanishes on exactly that sector where reduced Pfaffian is non-vanishing~\footnote{We will present the systematic study of reducing CHY formulas to four dimensions in~\cite{Zhang:2016rzb} . See also~\cite{Du:2016fwe} for a related study.}. In this sense ${\cal P}_n$ is strictly ``orthogonal" to the reduced Pfaffian, which means that the product of them vanishes in all sectors and cannot produce any non-zero amplitudes in four dimensions! This is the origin of the vanishing $R^2$ amplitude in four dimensions, which is the KLT of $F^3$ and Yang-Mills amplitudes.

Remarkably, we will also learn how self-dual and anti-self-dual parts appear from our formulas in four dimensions. We find that all the solution sectors that contributes to $F^3$ and $R^3$ amplitudes can be naturally divided into two complementary groups; $M^{F^3_+}_n$ and $M^{F^3_-}_n$ (similarly $M^{R^3_+}_n$ and $M^{R^3_-}_n$) are given by the sum of contributions from the two groups respectively, which also explains their orthogonality. In addition to providing a proof for \eqref{4drel}, our formulas show other nice features of $F^3$ amplitudes in four dimensions as well, such as the ``Parke-Taylor-like" formula for $M^{F^3_+}_n$ with three negative-helicity gluons~\cite{Broedel:2012rc}.

The paper is organized as follows. After briefly review the CHY formulas for Yang-Mills and gravity as well as KLT relations in section 2, we introduce the new ingredient ${\cal P}_n$ which lead to CHY formulas for all these amplitudes from higher-dimensional operators in section 3. In section 4, we discuss ${\cal P}_n$ in four dimensions, including its orthogonality to the reduced Pfaffian, and the split into self-dual and anti-self dual parts. Discussions and an appendix on reducing to four dimensions will be presented in the end.

\section{A Brief Review of CHY and KLT}

The universal part of CHY formulas contains the so-called scattering equations~\cite{Cachazo:2013iaa, Cachazo:2013gna, Cachazo:2013hca}
\be
 {\cal E}_a:=\sum_{b\neq a} \frac{s_{a\,b}}{\sigma_{a}-\sigma_{b}} \,=\, 0,
  \qquad \text{for}\quad a=1, 2, \ldots, n,
  \label{scatt}
\ee
where $s_{a\,b}=(k_a+k_b)^2=2 k_a\cdot k_b$, $\sigma_a$ is the $a^{\rm th}$ puncture. The tree-level S-matrix of $n$ massless particles is written as an integral localized on the support of \eqref{scatt}
\be\label{general}
M_n=\frac 1 {{\rm vol~SL}(2,\mathbb{C})}\,\int\,\prod_{a=1}^n d\,\sigma_a~\prod_{a=1}^n {}' \,\delta({\cal E}_a)~{\cal I}_n(\{\sigma, k, \ldots\})=\sum_{\rm solutions} \frac{{\cal I}_n(\{\sigma, k, \ldots\})}{J_n}\,,
\ee
where the precise definition of the integral measure including delta functions can be found in~\cite{Cachazo:2013hca}, and ${\cal I}_n$ is the CHY integrand that defines the theory. In the second equality one sums over $(n{-}3)!$ solutions of \eqref{scatt}, with $J_n$ the Jacobian of delta functions. In particular, the integrands for tree amplitudes in gravity, in Yang-Mills and a bi-adjoint $\phi^3$ theory are~\cite{Cachazo:2013iea}:
\be\label{integrands}
{\cal I}_n^{\rm GR}= {\rm Pf}' \bm{\Psi}_n (\epsilon)~{\rm Pf}' \bm{\Psi}_n (\tilde\epsilon)\,, \quad {\cal I}_n^{\rm YM}={\cal C}_n~{\rm Pf}' \bm{\Psi}_n\,,\qquad {\cal I}_n^{\phi^3}={\cal C}_n~\tilde{\cal C}_n\,.
\ee
The two ingredients are the Parke-Taylor factor which can be dressed with color factors,
\be
{\rm PT} (\alpha):=\frac 1 {\sigma_{\alpha(1), \alpha(2)} \cdots \sigma_{\alpha(n), \alpha(1)}}\,, \quad {\cal C}_n=\sum_{\alpha\in S_{n{-}1}}~{\rm Tr} (T^{I_\alpha(1)}~T^{I_\alpha(2)}\cdots T^{I_\alpha(n)})~{\rm PT}(\alpha)\,,
\ee
and a $2n\times 2n$ skew matrix $\Psi_n$ that depends on polarization vectors~\footnote{Here the polarization tensor for gravity is $\xi^{\mu\,\nu}=\epsilon^\mu \tilde\epsilon^{\nu}$, which is Einstein gravity coupled to a dilation and an anti-symmetric tensor, in the field theory limit of closed string theory.}:
\be
\small{\bm{\Psi}_n:=\left(
\begin{array}{cc}
A&-C^T\\
C&B
\end{array}
\right)\,;\quad 
A_{a,b}=\begin{cases}
\frac{k_a\cdot k_b}{\sigma_{a,b}}&a\neq b\\0&a=b
\end{cases},~
B_{a,b}=\begin{cases}
\frac{\epsilon_a\cdot\epsilon_b}{\sigma_{a,b}}&a\neq b\\0&a=b
\end{cases},~
C_{a,b}=\begin{cases}
\frac{\epsilon_a\cdot k_b}{\sigma_{a,b}}&a\neq b\\-\sum_{c\neq a}C_{a,c}&a=b
\end{cases},}
\ee
Note that the matrix is degenerate since it has two null vectors, but we can define its {\it reduced Pfaffian} by deleting two columns and rows among the first $n$:
\be
{\rm Pf}\,\bm{\Psi}_n=0\,;\qquad {\rm Pf}{}'\,\bm{\Psi}_n:=\frac{(-)^{a{+}b}}{\sigma_{a\,b}} {\rm Pf} |\Psi_n|^{a\,b}_{a\,b}\,,\quad {\rm with}\quad 1\leq a<b\leq n\,.
\ee
This definition is permutation invariant, and it has the appropriate $\SL2C$ weight and correct mass dimension, $[{\rm mass}]^{n{-}2}$ for producing Yang-Mills and gravity amplitudes via \eqref{general} and \eqref{integrands}. The most important property of ${\rm Pf}' \Psi_n$ is that on the support of scattering equations, it is invariant under gauge transformation $\epsilon^\mu_a \to \epsilon^\mu_a + \alpha k_a^\mu$~\cite{Cachazo:2013hca}

After color decomposition, one obtains color-ordered, partial amplitudes,  $M^{\rm YM} (\alpha)$, for Yang-Mills and double-partial amplitudes, $m (\alpha|\beta)$, for bi-adjoint scalar theory, with ${\rm PT}$ factors in the integrands. The field-theory limit of KLT relations can now be expressed as:
\be\label{KLT}
M^{\rm GR}_n =M_n^{\rm YM} \otimes_{\rm KLT} M_n^{\rm YM} :=\sum_{\alpha, \beta\in S_{n{-}3}}~M^{\rm YM} (\alpha)~m^{-1} (\alpha |\beta)~M^{\rm YM} (\beta)\,,
\ee
where $\alpha,\beta$ are in a basis of $(n-3)!$ orderings~\cite{Stieberger:2009hq, BjerrumBohr:2009rd}, and the KLT product of two sets of amplitudes is defined as their bilinear with the kernel given by the inverse of the matrix $(n{-}3)!\times (n{-}3)!$ matrix ${\bm m}$~\cite{Broedel:2013tta}. It is a simple linear-algebra proof \cite{Cachazo:2014xea} that \eqref{KLT} follows from \eqref{general} and \eqref{integrands}, which applies to general theories. Given any theory with CHY formula with its integrand of the form ${\cal I}_n^{\rm target}={\rm L}_n\,{\rm R}_n$, we can define two sets of partial amplitudes $M_n^{\rm L (R)}$ from CHY formula with integrands ${\cal I}_n^{\rm L (R)}={\rm PT}\,{\rm L}_n ({\rm R}_n)$ respectively. Then we have a general KLT relations among these amplitudes, $M^{\rm target}_n=M_n^{\rm L} \otimes_{\rm KLT} M_n^{\rm R}$.

Now we can write down the general form of the CHY formula for these amplitudes from higher-dimensional operators. Given that $F^3$ amplitudes have the same color-decomposition as well as BCJ relations as Yang-Mills amplitudes, one can always write its CHY integrand as the product of ${\cal C}_n$ (Parke-Taylor factor for partial amplitude) and a permutation invariant object that depends on polarizations. Let us call this new object as ${\cal P}_n (\epsilon)$, which must also be gauge invariant, and have mass dimension higher than ${\rm Pf}'\Psi$ by two. Now the KLT relations formulated above immediately imply CHY formulas for $M_n^{{\cal O}(\alpha'^2)}$, $M_n^{\cal O(\alpha')}$. The form of CHY integrands for these amplitudes are:
\be\label{newintegrands}
\boxed{{\cal I}_n^{F^3}={\cal C}_n \, {\cal P}_n (\epsilon)\,, \quad {\cal I}_n^{R^3}= {\cal P}_n (\epsilon)\, {\cal P}_n (\tilde\epsilon)\,,\quad {\cal I}_n^{R^2}={\cal P}_n (\epsilon)\, {\rm Pf}'\,\bm{\Psi} (\tilde\epsilon)\,.}
\ee
In the remainder of the paper, we will present the result for ${\cal P}_n$ and study its various interesting properties, such as soft limits and reduction to four dimensions.

\section{A New Ingredient in CHY Formulation}

In this section we will generalize ${\rm Pf}' \Psi_n$ to the new object ${\cal P}_n$. As basic requirements, it must be permutation and gauge invariant, must have the same $\SL2C$ weight and dimension $[{\rm mass}]^n$. The most natural and perfect candidate would of course be ${\rm Pf}\,{\bm \Psi}_n$ if it had not been zero! Nevertheless, we will see that ${\cal P}_n$ can be built from ${\rm Pf}\,\bm{\Psi}_n$. Let's  first give a natural decomposition of ${\rm Pf}\,\bm{\Psi}_n$ into objects that already satisfy all the conditions above individually. These will be the building blocks for our ${\cal P}_n$.

This interesting decomposition was essentially introduced in~\cite{Lam:2016tlk}. From the definition of Pfaffian and thanks to the special structure of $2n \times 2n$ matrix $\bm{\Psi}_n$, we can expand ${\rm Pf}\bm{\Psi}_n$ as a sum over $n!$ permutations of labels $1,2,\ldots, n$, denoted as $p\in S_n$
\ba\label{psi1}
{\rm Pf}\,\bm{\Psi}_n=\sum_{p\in S_n}\,{\rm sgn}(p)\,\Psi_p=\sum_{p\in S_n}\,{\rm sgn}(p)\,\Psi_I\Psi_J\cdots\Psi_K,
\ea
where ${\rm sgn}(p)$ denotes the signature of the permutation $p$ and in the second equality, we use the unique decomposition of any permutation $p$
into disjoint cycles $I,J,\cdots,K$ given by
\ba
I=(a_1a_2\cdots a_i),\quad J=(b_1b_2\cdots b_j),\cdots ,\quad K=(c_1c_2\cdots c_k)\;;
\ea
each $\Psi_p$ is the product of its ``cycle factors" $\Psi_I\Psi_J\cdots\Psi_K$, which we define now. When the length of a cycle equals one, its cycle factor $\Psi_{(a)}$ is given by the diagonal of $C$-matrix:
\ba\label{psii}
\Psi_{(a)}:=C_{aa}=-\sum_{b\neq a}\frac{\e_a\c k_b}{\s_{ab}}\;,
\ea
and when the length exceeds one {\it e.g.} $i>1$, the cycle factor is given by
\ba\label{psi2}
\Psi_I=\Psi_{(a_1a_2\cdots a_i)}:=\frac{\frac{1}{2}~\mathrm{tr}(f_{a_1}f_{a_2}\cdots f_{a_i})}{\s_{a_1a_2}\s_{a_2a_3}\cdots\s_{a_ia_1}}
\qquad \mathrm{with}\quad f_a^{\mu\nu}=k_a^{\mu}\e_a^{\nu}-\e_a^{\mu} k_a^{\nu}\;.
\ea
Here the trace is over Lorentz indices and $f^{\mu\nu}$ are the linearized field strengths of gluons. Note that the decomposition is manifestly gauge invariant: for cycle factors with length more than 1 \eqref{psi2}, the trace of linearized field strengths is gauge invariant, while for $1$-cycles, \eqref{psii}, the factor is gauge invariant on the support of scattering equations.

The proof for the decomposition is elementary and we refer to \cite{Lam:2016tlk} for more details. Let us look at some examples to illustrate the procedure. For ${\rm Pf}\bm{\Psi}_2$ we immediately have
\ba
{\rm Pf}\bm{\Psi}_2=&&C_{11}C_{22}+(A_{12}B_{21}- C_{12}C_{21})=\Psi_{(1)}\Psi_{(2)}\,-\,\Psi_{(12)}\;.
\ea
In the second equality, $C_{11}C_{22}=\Psi_{(1)}\Psi_{(2)}$ corresponds to the permutation $(1) (2)$; the terms $A_{12}B_{21}$ and $C_{12}C_{21}$ have a common denominator  $\s_{12}\s_{21}$, and the numerators combine to  $k_1\c k_2 \,\e_2\c \e_1- \e_1\c k_2\, \e_2\c k_1=\frac{1}{2}\mathrm{tr}(f_{1}f_{2})$, thus we have the desired cycle factor $\Psi_{(12)}$.

For  ${\rm Pf}\Psi_3$, there are new building blocks of length $3$, $(123)$ and $(321)$. Four terms from the expansion of ${\rm Pf}\Psi_3$ corresponds to $(123)$ with a common denominator $\s_{12}\s_{23}\s_{31}$, \ba
(123):\quad C_{12}C_{23}C_{31}-C_{12}A_{23}B_{31}-C_{23}A_{31}B_{12}-C_{31}A_{12}B_{23}\;,
\ea
and similarly for $(321)$ (with denominator $\s_{32}\s_{21}\s_{13}=-\s_{12}\s_{23} \s_{31}$). Note that neither of them is gauge invariant, but the sum of the two is: the eight terms of their numerators nicely combine to $\mathrm{tr}(f_{1}f_{2}f_{3})$! It is convenient to assign to each of them half of the trace, {\it i.e.} $\frac{1}{2}\mathrm{tr}(f_{1}f_{2}f_{3})$, as the numerator. Thus we arrive at \eqref{psi2} as expected, and we have
\ba\label{ps3}
{\rm Pf}\bm{\Psi}_3=\Psi_{(1)}\Psi_{(2)}\Psi_{(3)}-(\Psi_{(1)}\Psi_{(23)}
+\Psi_{(2)}\Psi_{(13)}+\Psi_{(3)}\Psi_{(12)})+\Psi_{(123)}+\Psi_{(321)}\;.
\ea

Given the decomposition of ${\rm Pf}\bm{\Psi}_n$ as in \eqref{psi1}, we can classify $\Psi_p$'s by the lengths of its cycles $\{i,j,..., k\}$. For example, in \eqref{ps3} the first term is of the type $\{1,1,1\}$ as it is the product of three 1-cycles; the three terms in the bracket are all of the type $\{1,2\}$, and the last two terms are $\{3\}$. The reason for doing so is of course to group together terms in \eqref{psi1} of the same type, and write a manifestly permutation invariant decomposition of ${\rm Pf}\bm{\Psi}_n$. Furthermore, note that the signature of a permutation is given by $n$ minus the number of cycles, so one can sum over all permutations of the same type with identical signs. Let's define {\it permutation invariant} building blocks as follows:
\ba\label{P}
P_{i_1\,i_2\,\cdots\,i_r}
:=\sum_{\tiny{\substack{|I_1|=i_1,|I_2|=i_2,\cdots,|I_r|=i_r}}}\Psi_{I_1}\Psi_{I_2}\cdots\Psi_{I_r}\;,
\ea
which is a sum of $\Psi_p$'s for all permutations of the same type $\{i_1, i_2,\ldots, i_r\}$, with
\ba
 i_1+ i_2+\cdots +i_r=n\;, \quad {\rm and~the~convention:}\quad  i_1\leq i_2\cdots \leq i_r
\ea
Here each $P$ is by construction permutation invariant. Let's again see some examples:
\ba\label{pex}
P_{11\cdots 1}=&&\Psi_{(1)}\Psi_{(2)}\cdots \Psi_{ (n)}=C_{11}C_{22}\cdots C_{nn}\;,\nl
P_{12}\;=&&\Psi_{(1)}\Psi_{(23)}+\Psi_{(2)}\Psi_{(13)}+\Psi_{(3)}\Psi_{(12)}\;,\nl
P_{22}\;=&&\Psi_{(12)}\Psi_{(34)}+\Psi_{(13)}\Psi_{(24)}+\Psi_{(14)}\Psi_{(23)}\;,\nl
P_{13}\;=&&\Psi_{(1)}\Psi_{(234)}+\Psi_{(2)}\Psi_{(134)}+\Psi_{(3)}\Psi_{(124)}+\Psi_{(4)}\Psi_{(123)}
\nl
&&+\Psi_{(1)}\Psi_{(432)}+\Psi_{(2)}\Psi_{(431)}+\Psi_{(3)}\Psi_{(421)}+\Psi_{(4)}\Psi_{(321)}\;.
\ea
With \eqref{P},  \eqref{psi1} can be rewritten as a permutation invariant decomposition of ${\rm Pf}\bm{\Psi}_n$:
\ba\label{cp}
{\rm Pf}\bm{\Psi}_n=\displaystyle \sum\limits_{\tiny \substack{1\leq i_1\leq i_2\leq\cdots\leq i_m\leq n \\i_1+i_2+\cdots+i_m=n}}
~(-)^{n-m}\, P_{i_1i_2\cdots i_m}\;.
\ea
Let us spell out the decomposition for $n=3$ (see \eqref{ps3}) and $n=4,5$:
\ba
{\rm Pf}\bm{\Psi}_3=&&P_{111}-P_{12}+P_{3}\;.\nl
{\rm Pf}\bm{\Psi}_4=&&P_{1111}-P_{112}+P_{13}+P_{22}-P_{4}\;.\nl
{\rm Pf}\bm{\Psi}_5=&&P_{11111}-P_{1112}+P_{113}+P_{122}-P_{14}-P_{23}+P_{5}\;.
\ea
Note that each $P$ and thus any linear combination of them immediately satisfy our conditions above: correct $\SL2C$ weight and mass dimension, permutation and gauge invariance. ${\rm Pf}\bm{\Psi}_n=0$ means that these building blocks are not all independent: \eqref{cp} gives linear relations between different $P$'s. In the following, we will present a very special linear combination that leads to the correct CHY formula $F^3$ amplitudes. Let us first present the answer and then discuss its special properties.

It turns out one only needs to modify coefficients of \eqref{cp} a bit to obtain ${\cal P}_n$:

\ba\label{2}
\boxed{\mathcal{P}_n=\displaystyle \sum\limits_{\tiny \substack{1\leq i_1\leq i_2\leq\cdots\leq i_m\leq n \\i_1+i_2+\cdots+i_m=n}}~(-)^{n-m}~\left( N_{i>1} + c\right)~P_{i_1\,i_2\,\cdots\,i_m}\;,}
\ea
were $N_{i>1}$ denotes the number of indices in $i_1, i_2, \cdots, i_m$ which are larger than $1$, or the number of cycles with length at least $2$; $c$ is just any constant because we can add any multiple of \eqref{cp} without changing the answer. The formula can simplify when we choose the constant $c$ to be certain integers. For example, two convenient choices are $c=-1$ and $c=0$ respectively, and we have:
\ba\label{P345}
\begin{aligned}
{\cal P}_3&=-P_{111}&=&-P_{12}+P_{3}\\
{\cal P}_4&=-P_{1111}+P_{22}&=&-P_{112}+P_{13}+2P_{22}-P_{4}\\
{\cal P}_5&=-P_{11111}+P_{122}-P_{23}&=&-P_{1112}+P_{113}+2P_{122}-P_{14}-2P_{23}+P_{5}
\end{aligned}
\ea
where in the first representation $P_{11\cdots 1}$ is always present with $-1$ but any $P$ with only one index $i>1$ are always absent (including $P_n$); in the second one $P_{11\cdots1}$ is always absent. To give one more example, here is ${\cal P}_6$ with $c=-1$:
\ba\label{P6}
{\cal P}_6=&&-P_{111111}+P_{1122}-P_{123}-2P_{222}+P_{24}+P_{33}\;.
\ea

We have checked thoroughly that \eqref{2} gives correct $F^3$ amplitudes. First of all, one can easily verify that for $n=3,4$, the formula reproduces correct amplitudes as computed from Feynman diagrams. In the next section we will provide very strong evidence for its validity, including checks for all helicities up to $n=8$ and some all-multiplicity results. Here we provide another important check, that is its behavior under soft limits.

Recall that in CHY formula for gravity and Yang-Mils, Weinberg's soft theorem~\cite{Weinberg:1965nx} becomes manifest due to the simple soft limits of ${\rm Pf}' \Psi_n$. Let us take the $a$-th particle to be soft, that is $k_a^{\mu}=\tau q^{\mu}$ with $\tau\rightarrow 0$. The soft graviton and soft gluon theorems are guaranteed by CHY formula in as long as we have ${\rm Pf}' \Psi_n \to C_{a a} {\rm Pf}'\Psi_{n{-}1} + {\cal O}(\tau)$ where ${\rm Pf}'\Psi_{n{-}1}$ has only hard particles. What is important here is that soft theorems are universal, thus apply to amplitudes from higher-dimensional operators as well~\cite{Bianchi:2014gla}. For this to work the soft behavior of ${\cal P}_n$ must be identical to that of ${\rm Pf}' \Psi_n$. Let's check this explicitly.

Note that in any term of ${\cal P}_n$, $a$ must be in one of the cycles, and there are two possibilities. If it is a cycle of length at least $2$, then in the numerator $f_a^{\mu\nu} \to {\cal O}(\tau)$ as $\tau \to 0$ while the denominator remains finite, thus the cycle factor vanishes, $\Psi_{(\ldots a)} \to {\cal O}(\tau)$. On the other hand, if it is in an 1-cycle then it remains finite
\ba\label{soft2}
\Psi_{(a)}=C_{a a}=-\sum_{b\neq a} \frac{\epsilon_a\cdot k_{b}}{\sigma_{a\,b}}\to {\cal O}(1)\;.
\ea
For any $P_{i_1 i_2 \cdots i_m}$ with lengths of all cycles being at least $2$, i.e. $i_m\geq i_{m-1}\geq \ldots \geq i_1>1$ (note that here $N_{i>1}=m$), it vanishes as ${\cal O}(\tau)$ in any single soft limit. Therefore requiring ${\cal P}_n$ to have correct soft behavior {\it cannot} constrain coefficients of such terms at all.

On the contrary, the soft limit puts very strong constraints on those $P$'s that have at least one cycle with length $1$, i.e. $i_1=1$. In the $k_a\to 0$ limit only those terms with $a$ in an 1-cycle survive and dominate in the limit (other terms still vanish). Thus for any single soft limit,  $P_{1, i_2, \ldots, i_m} \to C_{a a} P_{i_2, \ldots, i_m}$ where $P_{i_2, \ldots, i_m}$ is the $(n-1)$-point building block with particle $a$ removed. Note that in \eqref{2} the coefficients are determined by $N_{i>1}$ and independent of 1-cycles, thus $P_{1, i_2, \ldots, i_m}$ and $P_{i_2, \ldots, i_m}$ have exactly identical coefficients. Thus we see that \eqref{2} indeed satisfies (${\cal P}_{n{-}1}$ contains all particles except $a$):
\ba
\mathcal{P}_n\;\overset{k_a\rightarrow 0}{\longrightarrow}\; C_{aa}~\mathcal{P}_{n-1}\;.
\ea

In other words, ${\cal P}_n$ splits into two parts that behave very differently under soft limit
\be
{\cal P}_n=\left(\sum_{1=i_1\leq i_2 \leq \ldots \leq i_m } + \sum_{1<i_1 \leq  i_2 \leq \ldots \leq i_m}\right) ~(-)^{n-m}~\left( N_{i>1} + c\right)~P_{i_1, i_2,\ldots,i_m}\,,
\ee
where the first part is essentially fixed by soft limit, namely the coefficients for any $P_{1, \ldots}$ must be the same as that for the $P$ with $1$ removed. This explains why the coefficient should not depend on how many 1-cycle are there. However, we have seen that soft limits put no constraints on the coefficients of the second part, which vanishes term by term.

We believe that correct behavior under factorization limits of ${\cal P}_n$ can completely fix the coefficients of the second part. However, even without resorting to that, we will now show that the coefficients in \eqref{2} are strongly constrained by another remarkable property of ${\cal P}_n$ in four dimensions, namely it is orthogonal to ${\rm Pf}' {\bm \Psi}_n$.

\section{Four Dimensions, Orthogonality and Self-Duality}

In this section we study important properties of ${\cal P}_n$ in four dimensions. Details for the reduction to four dimensions will be presented in~\cite{Zhang:2016rzb}. As discussed in~\cite{Cachazo:2013iaa,He:2016vfi} and briefly reviewed in appendix A, in four dimensions, the $(n{-}3)!$ solutions of scattering equations fall into $n-3$ sectors labeled by $k'=2,3,\ldots,n-2$, thus \eqref{general} becomes a sum over sectors and we define the contribution from sector $k'$ as $T^{(k')}$:
\be\label{decompose}
M_n=\sum_{\rm solutions}~\frac{{\cal I}_n}{J_n}=\sum_{k'=2}^{n{-}2} \left( \sum_{k'~{\rm sec. sol.}}~\frac{{\cal I}_n}{J_n}\right):=\sum_{k'=2}^{n{-}2}~T^{(k')}_n\,
\ee
On the other hand, for massless particles with spin in four dimensions, we specify helicities of the $n$ particles, which can be divided into a set of particles with negative helicities, $-$, and the complementary one $+$. We denote the helicity sectors by the number of negative-helicity particles, $k:=|-|=0,1,\ldots, n$ ($|+|=n-k$) and call the helicity amplitude in this sector, $M_{n,k}$.  {\it A priori} there is no relation between solution sector and helicity sector.

However, it is known that ${\rm Pf}' \Psi_n$ vanishes unless $k=k'$ (in particular it vanishes for $k=0,1,n{-}1,n$), which means Yang-Mills and gravity amplitudes in helicity sector $k$ only receives contribution from solutions in sector $k'=k$
\be\label{YMGR}
T_{n,k}^{(k')}=0\,, \quad {\rm for~any}~k'\neq k\,,\quad \Rightarrow\quad M_{n,k}=T_{n,k}^{(k'=k)}\,,\quad {\rm for~YM,~GR,~etc.}
\ee
Now we show that exactly the opposite is true for ${\cal P}_n$, namely it vanishes for $k'=k$, thus
\be\label{F3R3}
\boxed{T_{n,k}^{(k'=k)}=0\,,\quad \Rightarrow\quad M_{n,k}^{R^2}=0 \quad \& \quad M_{n,k}=\sum_{k'\neq k} T_{n,k}^{(k')} \qquad {\rm for}~F^3,~R^3,~{\rm etc.}}
\ee

The starting point of the reduction is the simple reduction of trace of linearized field strengths in four dimensions for any assignment of helicities:
\be\label{tr}
\mathrm{tr}~(f_{a_1}\,f_{a_2} \cdots  f_{a_x})=\begin{cases}
2~\langle a_1a_2\rangle~\langle a_2 a_3\rangle \cdots \langle a_x a_1\rangle\,, \qquad &\{a_1,a_2,\cdots a_x\}\subset -\\
2~[a_x a_{x-1}]~[a_{x-1}a_{x-2}] \cdots [a_1 a_x]\,, \qquad &\{a_1,a_2,\cdots a_x\}\subset +\\
\langle b_1b_2\rangle \cdots \langle b_y b_1\rangle
 [p_{z} p_{z-1}]\cdots [p_1 p_z]
\,,& \quad {\rm otherwise}
\end{cases}
\;,
\ee
Here $b_1,b_2,\cdots,b_y$ are all the particles of negative helicity from $a_1,a_2,\cdots,a_x$ with its ordering unchanged and similarly $p_1,p_2,\cdots,p_z$ are all the particles of positive helicity from $a_1,a_2,\cdots,a_x$ with its ordering unchanged. Note that $\mathrm{tr}~(f_{a_1}\,f_{a_2} \cdots  f_{a_x})$  directly vanishes if there is only  one particle of negative helicity or  only one particle of positive helicity in $a_1,a_2,\cdots,a_x$. However we see that the remaining case still effectively vanish as we always add up all permutations (see \eqref{P}) while
\ba
\sum_{\{\a\}\in {\rm OP}(\{b_1,b_2,\cdots,b_y\},\{c_1,c_2,\cdots,c_z\})}\frac{1}{\s_{(\{\a\})}}=0\;.
\ea
Here the sum is over ordered permutations $``$OP$"$, namely permutations of the labels in the joined set $\{b_1,b_2,\cdots,b_y\},\{c_1,c_2,\cdots,c_z\}$
such that the ordering within $\{b_1,b_2,\cdots,b_y\}$ and $\{c_1,c_2,\cdots,c_z\}$ is preserved. Therefore, in the sum of \eqref{psi1}, we can effectively write $\mathrm{tr}~(f_{a_1}\,f_{a_2} \cdots f_{a_x})$ in 4d in a remarkably simple way:
\be\label{tr}
\frac 1 2 \mathrm{tr}~(f_{a_1}\,f_{a_2} \cdots f_{a_x}) \to \begin{cases}
\langle a_1a_2\rangle~\langle a_2 a_3\rangle \cdots \langle a_x a_1\rangle\,, \qquad &\{a_1,a_2,\cdots a_x\}\subset -\\
[a_1 a_2]~[a_2a_3] \cdots [a_x a_1]\,, \qquad &\{a_1,a_2,\cdots a_x\}\subset +\\
0\,, \qquad &{\rm otherwise}
\end{cases}
\;,
\ee

Motivated by \eqref{tr}, we recall the off-diagonal elements of the $k\times k$ matrix $\bm{h}_k$ and $(n{-}k)\times (n{-}k)$ one $\bm{\tilde{h}}_{n-k}$ essentially introduced in~\cite{Geyer:2014fka} (see also \cite{Cachazo:2012kg,Cachazo:2012pz}):
\ba
h_{ab}=\frac{\<ab\>}{\s_{ab}} \quad a\neq b,\;a,b\in -\;,
\qquad
\tilde{h}_{ab}=\frac{[ab]}{\s_{ab}}\ \ \quad a\neq b,\;a,b\in +\;.
\ea
As discussed above, it is clear that when we have any cycle factor with length at least 2, effectively it reduces to the chain product of such off-diagonal elements in 4d:
\ba\label{psiI}
\Psi_{(a_1a_2\cdots a_x)} \to 
\begin{cases}
h_{a_1a_2}h_{a_2a_3}\cdots h_{a_x a_1} \qquad &\{a_1,a_2,\cdots a_x\}\subset -\\
\tilde{h}_{a_1a_2}\tilde{h}_{a_2a_3}\cdots \tilde{h}_{a_x a_1} \qquad &\{a_1,a_2,\cdots a_x\}\subset +\\
0 &{\rm otherwise}
\end{cases}
\;,
\ea
To this point we have not used scattering equations or solution sectors in four dimensions. As we prove in the appendix A, the really non-trivial part of the reduction concerns 1-cycle, or the diagonal entries of $C$-matrix. Note that $\Psi_{(a)}=C_{aa}$ is only gauge invariant on the support of scattering equations, so it is not surprising that to reduce it nicely one needs to use scattering equations in four dimensions.

We first discuss the $k'=k$ case: miraculously, by plugging in scattering equations in $k'=k$ sector, $C_{aa}$ reduces to diagonal entries of $\bm{h}_k$ or $\bm{\tilde{h}}_{n-k}$~\cite{Geyer:2014fka} depending on the helicity:
\ba\label{diag}
h_{aa}=C_{aa}^-=-\sum_{\tiny{\substack{b\neq a\\b\in -}}}\frac{t_b}{t_a}\frac{\<ab\>}{\s_{ab}}\quad a\in-;,
\qquad
\tilde{h}_{aa}=C_{aa}^+=-\sum_{\tiny{\substack{b\neq a\\b\in +}}}\frac{t_b}{t_a}\frac{[ab]}{\s_{ab}}\quad a\in+\;.
\ea
The details of the proof is given in appendix A;  $t$'s and $\tilde t$'s are determined by scattering equations in 4d but here we can just view them as free variables, and the important thing is that each diagonal entry is a linear combination of off-diagonal entries in that row/column. With diagonal entries of $\bm{h}_k$ or $\bm{\tilde{h}}_{n-k}$ defined as \eqref{diag}, now the reduction for $\Psi_{(a_1a_2\cdots a_x)}$ with $x>1$ or $x=1$ (for $k'=k$) are unified in one nice formula, \eqref{psiI}.

Before we prove the vanishing of ${\cal P}_n$ in $k'=k$ sector, let us again return to our favorite ${\rm Pf}{\bm \Psi}_n$, and first show the following identity as a warm up:
\ba\label{id1}
{\rm Pf}{\bm \Psi}_n=\mathrm{det}\,{\bm h}_{k}\,\mathrm{det}\,{\bm {\tilde{h}}}_{n-k}\;.
\ea
Obviously both $\mathrm{det}\,{\bm h}_{k}$ and $\mathrm{det}\,{\bm {\tilde{h}}}_{n-k}$ vanish since they both have a null vector; this is consistent with the fact that
${\rm Pf}{\bm \Psi}_n$ vanishes due to the two null vectors. To show \eqref{id1}, we decompose $\mathrm{det}\,{\bm h}_{k}$, $\mathrm{det}\,{\bm {\tilde{h}}}_{n-k}$ in a way similar to that of ${\rm Pf}{\bm \Psi}_n$, {\it e.g.} for $\mathrm{det}\,{\bm h}_{k}$ we have
\ba
\mathrm{det}\;\bm{h}_k=\sum_{q\in S_k}{\rm sgn}(q) h_{I_1}h_{I_2}\cdots h_{I_s}\;,\quad {\rm with}\  h_{I}=h_{(a_1a_2\cdots a_i)}=h_{a_1a_2}h_{a_2a_3}\cdots h_{a_ia_1}\;,
\ea
where the sum is over all permutations of particles of negative helicity, {\it i.e.} $q\in S_k$ and  $I_1,I_2,\cdots,I_s$ are the cycles of the permutation $q$. We can further define
\ba\label{H}
H_{i_1i_2\cdots i_{\ell}}
=\sum_{\tiny{\substack{|I_1|=i_1,|I_2|=i_2,\cdots,|I_t|=i_{\ell}}}}h_{I_1}h_{I_2}\cdots h_{I_{\ell}}\;
\qquad \mathrm{with}\; i_1\leq i_2\cdots \leq i_{\ell}\;,
\ea
then $\mathrm{det}\;\bm{h}_k$ can be rewritten as a sum of $H$ and similarly works $\mathrm{det}\; \bm{\tilde{h}}_{n-k}$,
\ba\label{Hdet}
\mathrm{det}\;\bm{h}_k=\sum_{\{i\}_k^{\ell}}(-)^{k-{\ell}}H_{i_1i_2\cdots i_{\ell}}\;,
\qquad
\mathrm{det}\; \bm{\tilde{h}}_{n-k}=\sum_{\{{\tilde i}\}_{n-k}^{{\tilde \ell}}}(-)^{n-k-{{\tilde \ell}}}\tilde{H}_{{\tilde i}_1{\tilde i}_2\cdots {\tilde i}_{{\tilde \ell}}}\;,
\ea
where we have introduced shorthand notation for the summation range,  $\{i\}_k^{\ell}$ means $i_1+ i_2 + \ldots i_{\ell}=k$ and $i_1 \leq i_2 \leq \cdots \leq i_{\ell}$ and similarly for $\{{\tilde i}\}_{n-k}^{{\tilde \ell}}$.

Both ${\cal P}_n$ and ${\rm Pf}\Psi_n$ are built from $P$'s, so the key identity here is for the reduction of the $P$'s, which nicely follow from \eqref{P}, \eqref{H} and \eqref{psiI}:
\ba\label{ph}
P_{i_1i_2\cdots i_m}=\sum H_{j_1j_2\cdots j_{\ell}}\tilde{H}_{\tilde{j}_1\tilde{j}_2\cdots\tilde{j}_{\tilde{{\ell}}}}\;,
\ea
where, recall that any cycle factor in $P$ is only non-vanishing when all particles belong to the same helicity set, thus the sum in $P$ ``factorizes" into sums in $-$ set and those in $+$ set, which give $H$ and $\tilde{H}$; the additional sum in \eqref{ph} is over all distinct partition of $i_1\,i_2\,\cdots\,i_m$ into two parts $j_1\,j_2\,\cdots\,j_{\ell}$ and $\tilde{j}_1\,\tilde{j}_2\,\cdots\,\tilde{j}_{\tilde{{\ell}}}$, with $j_1+ j_2+\cdots+ j_{\ell}=k$ and $\tilde{j}_1+\tilde{j}_2+\cdots+\tilde{j}_{\tilde{{\ell}}}=n-k$. For example, any $P_{11\cdots 1}$ reduces to $H_{11\cdots1}$ and  $\tilde{H}_{11\cdots1}$:
\ba
P_{\underbrace{11\cdots 1}_n} \overset{k}{\longrightarrow}  H_{\underbrace{11\cdots1}_{k}} \tilde{H}_{\underbrace{11\cdots1}_{n-k}}\;.
\ea
We have more examples for $n=4, k=2$ and $n=7,k=3$,
\ba
P_{22}\overset{k=2}{\longrightarrow} H_2\tilde{H}_2\;,\qquad  P_{13}\overset{k=2}{\longrightarrow} 0\;;\qquad \qquad P_{1123}\overset{k=3}{\longrightarrow} H_{12}\tilde{H}_{13}+H_{3}\tilde{H}_{112}\;.
\ea
Given \eqref{ph}, it is trivial to show \eqref{id1} using \eqref{cp} and \eqref{Hdet}. Although both sides vanish, this is still an example of the remarkable simplifications in a given sector in four dimensions: we see that most of the terms vanish and the number of terms are reduced from $n!$ to $k! \times (n{-}k)!$. Along the same line but in a more non-trivial way, similar simplification happens for the reduction of ${\rm Pf}'\Psi_n$ which will be present in~\cite{Zhang:2016rzb}.

We turn to the reduction of ${\cal P}_n$. By dividing $N_{i>1}$ in \eqref{2} into two parts $N_{j>1}$ and $N_{\tilde{j}>1}$ (set $c=0$) which depend on $-$ and $+$ sets respectively, and ${\cal P}_n$ reduces to:
\ba\label{pn1}
\left(\sum_{\{i\}_k^{\ell}}(-)^{k{-}{\ell}}\,N_{i{>}1}\,H_{i_1i_2\cdots i_{\ell}}\right) \mathrm{det}\,{\bm {\tilde{h}}}_{n{-}k}\;+\mathrm{det}\,{\bm h}_{k}\;\left(\sum_{ \{\tilde{i}\}_{n{-}k}^{\tilde{{\ell}}}}(-)^{n{-}k{-}\tilde{{\ell}}}\,N_{\tilde{i}>1}\,
\tilde{H}_{\tilde{i}_1\tilde{i}_2\cdots \tilde{i}_{\tilde{{\ell}}}}\right).
\ea
Thanks to the vanishing of $\mathrm{det}\,{\bm {\tilde{h}}}_{n-k}$ and $\mathrm{det}\,{\bm h}_{k}$, we immediately see that $\mathcal{P}_n$ vanishes for $k=k'$. Before proceeding, let's provide a few explicit examples of \eqref{pn1} for ${\cal P}_4$ and  ${\cal P}_5$:
\ba
{\cal P}_4
&\overset{k=2}{\longrightarrow}&-H_{11}\tilde{H}_{2}-H_{2}\tilde{H}_{11}
+2H_{2}\tilde{H}_{2}\nl
&=&-H_{2}(\tilde{H}_{11}-\tilde{H}_{2})+({H}_{11}-{H}_{2})(-\tilde{H}_{2})\nl
&=&-H_{2}\;\mathrm{det}\,{\bm {\tilde{h}}}_{2}+\mathrm{det}\,{\bm h}_2\,(-\tilde{H}_{2})\;,
\ea

\ba\label{example5}
{\cal P}_5
&\overset{k=3}{\longrightarrow}&-H_{111}\tilde{H}_{2}-H_{12}\tilde{H}_{11}+H_{3}\tilde{H}_{11}
+2H_{12}\tilde{H}_{2}-2H_{3}\tilde{H}_{2}\nl
&=&(-H_{12}+H_{3})(\tilde{H}_{11}-\tilde{H}_{2})+({H}_{111}-{H}_{12}+{H}_{3})(-\tilde{H}_{2})\nl
&=&(-H_{12}+H_{3})\;\mathrm{det}\,{\bm {\tilde{h}}}_{3}+\mathrm{det}\,{\bm h}_2\,(-\tilde{H}_{2})\;,
\ea

It is a remarkable fact that ${\cal P}_n$ vanishes for $k'=k$ sector. As we mentioned before, this property can be used to constrain the second part of ${\cal P}_n$ which are not constrained by soft limits at all. Up to $n=8$, we found that the constraints that ${\cal P}_n$ vanishes for $k'=k$ sector for all helicity sectors uniquely fix all coefficients in ${\cal P}_n$.

This property means that ${\cal P}_n$ is completely orthogonal to ${\rm Pf}'\Psi_n$ in four dimensions. For evaluating helicity amplitudes for Yang-Mills/gravity vs. those for $F^3$ or $R^3$, one always uses complementary set of solutions of scattering equations. This seems to be the scattering-equation origin of the vanishing of 4d $R^2$ amplitudes, which has a CHY integrand ${\cal P}_n\,{\rm Pf}'\Psi_n$. In general dimensions, the integrand is of course non-zero, but once we reduce to four dimensions, it vanishes for every solution of scattering equations!

The derivation of \eqref{pn1} applies to any $k'\neq k$ case as well, with the only difference being that the reduction of $1$-cycle {\it i.e.} $C_{aa}$ needs to be modified. As shown in appendix A, we can generalize the diagonal entries of the two matrices ${\bm h}_{k}^{k'}$ and ${\bm {\tilde{h}}}_{n-k}^{k'}$ depending on the solution sector $k'$ and helicity sector $k$. The upshot is that \eqref{pn1} still holds for any $k' \neq k$ sector with generalized matrices ${\bm h}_{k}^{k'}$ and ${\bm {\tilde{h}}}_{n-k}^{k'}$. Just by inspecting the matrices, it turns out that we again have $\mathrm{det}\,{\bm {h}^{k'}_k}=0$ for $k'<k$ and $\mathrm{det}\,{\bm {\tilde{h}}^{k'}_{n-k}}=0$ for $k'>k$, thus for any $k'$, only one of the two terms in \eqref{pn1} remain non-vanishing.

In view of this, it becomes very natural to divide the sectors into two groups: those with $k'<k$ and those with $k'>k$, and the question is does this separation means anything sensible for $F^3$ and $R^3$ amplitudes in four dimensions? The answer is affirmative: the sum of contributions from the two complementary groups correspond to self-dual and anti-self-dual amplitudes, respectively. Let's write down this proposal for $F^3$ amplitudes:
\be\label{SD}
\boxed{M_{n,k}^{F_+^3}=\sum_{2\leq k'<k} T_{n,k}^{(k')}\,,\qquad M_{n,k}^{F_-^3}=\sum_{n-2\geq k'>k} T_{n,k}^{(k')}\,,\qquad 
{\rm for~}k=0,1,\ldots,n\,.}
\ee
An immediate consequence of \eqref{SD} is that self-dual and anti-self-dual parts are orthogonal, which implies the second part of the \eqref{4drel}. These are very non-trivial relations from usual representation of the amplitudes, but become totally obvious from \eqref{SD}. Given that their KLT vanishes, it immediately follows that \eqref{SD} also applies to $R^3$ amplitude.

There is very strong evidence that \eqref{SD} must be correct. First of all, it implies the well-known fact that for $k=0,1,2$, self-dual amplitudes vanishes (no $k'<2$) and there are only anti-self-dual amplitudes, while for $k=n,n{-}1,n{-}2$ there are only self-dual amplitudes (no $k'>n{-}2$). To provide more non-trivial evidence for \eqref{SD}, we have checked our proposal for self-dual, $F_+^3$ amplitudes against \cite{Broedel:2012rc} for all helicities up to eight points. We have evaluated our formula numerically for solutions in all sectors of $k'\neq k$, and find that the self-dual amplitude is the sum of those sectors listed in Table~\ref{ta1}.

\begin{table}[!htbp]
\centering
\begin{tabular}
{|p{0.8cm}|p{1cm}|p{1cm}|p{1cm}|p{1cm}|*{2}{p{1.3cm}|}}
  \hline  
  \diagbox{$n$}{ $k'$}{$k$} & 3 & 4 & 5 & 6 &7&8 \\
  \hline
 3&  2 & - & - & - & - & -  \\
   \hline
 4& 2 & 2,3 &- & -& - &-  \\
   \hline
5 & 2 & 2,3 & 2,3 & - & - & - \\
  \hline
 6 & 2 & 2,3 & 2,3,4 & 2,3,4 &- &- \\
   \hline
 7 & 2 & 2,3 & 2,3,4 & 2,3,4,5 & 2,3,4,5&-  \\
   \hline
8 & 2 & 2,3 & 2,3,4 &2,3,4,5 & 2,3,4,5,6 & 2,3,4,5,6 \\
  \hline
\end{tabular}
\caption{\label{ta1} {\large {Sectors contribute to the {\it self-dual} amplitudes $M^{F_+^3}_{n,k}$ and $M^{R_+^3}_{n,k}$ }}}
\end{table}

Our proposal suggests that there is a natural origin for self-dual and anti-self-dual amplitudes from solution sectors of scattering equations in 4d. Note that individually $T_{n,k}^{(k')}$ are not physical for general $k$ and $k'$, since they can contain spurious poles, as is familiar from the reduction of bi-adjoint $\phi^3$ to four dimensions~\cite{Cachazo:2016sdc}. The interesting thing is that unlike in the scalar case where one has to sum over all sectors, here by summing over subsets of sectors, namely those with $k'<k$ and those with $k'>k$, we already obtain physical amplitudes, $M_{n,k}^{F_+^3}$ and $M_{n,k}^{F_-^3}$, respectively.

There is a special case when we do not need to sum over sectors at all, and it also serves as an important check of the proposal \eqref{SD}. This is the $F^3_-$ amplitudes with three negative-helicity gluons, {\it i.e.} $k=3$, which receives the contribution only from $k'=2$ sector, $M_{n,3}^{F^3_+}=T_{n,3}^{(2)}$; moreover it is well known that there is a just a unique solution in that sector.

To be concrete, let's choose the three particles of negative helicity as $p,q,r$. Note that for $k=3$ and $k'=2$ the generalized version of \eqref{pn1} has the first term vanishes and the second term evaluates to (the details are given in appendix A):
\ba
{\cal P}_n=h_{pp}\,h_{qq}\,h_{rr}~\mathrm{det}\,{\bm {\tilde{h}}}_{n-3}^{2}\; \qquad {\rm when}\ k=3,k'=2\;.
\ea
We can choose to generalize $h_{rr}$ then from \eqref{haakp} we can directly obtain $h_{pp}h_{qq}=\langle p\,q\rangle^2/\sigma_{p\,q}^2$ and $h_{rr}=t_r^2\frac{t_p t_q \s_{pq}\<pq\>}{\s_{rp}^2\s_{rq}^2}$. As explained in appendix A, the seemingly complicated factor $\mathrm{det}\,{\bm {\tilde{h}}}_{n-3}^{2}$ is in fact canceled by a Jacobian factor ${\mathrm{det'}\,{\bm{h}}_{2}\mathrm{det'}\,{\bm{\tilde{h}}}_{n-2}}$ from the measure. Collecting all ingredients and evaluating on the unique solution of $k'=2$ sector \eqref{MHVsol}, we obtain the remarkably simple ``Parke-Taylor-like" formula for $F^3_+$ amplitude
\ba
M_{n,3}^{{ F}^3_+} (p^-, q^-, r^-) = \frac{(\<pq\>\<pr\>\<rp\>)^2}{\<12\>\<23\>\cdots\<n1\>}\;.
\ea
\section{Discussions}

In this paper we studied tree-level amplitudes from higher-dimensional operators, including the $F^3$ modification to Yang-Mills action, and those to Einstein gravity from bosonic closed strings at lowest orders.  We proposed new CHY formulas for these amplitudes, \eqref{newintegrands}, and all the modifications are naturally encoded in one new ingredient, ${\cal P}_n$ as given in \eqref{2}. The reduced Pfaffian is the natural object for Yang-Mills and gravity amplitudes, and ${\cal P}_n$ is the first genuinely new object that generalizes it for higher-dimensional operators. By construction it is manifestly permutation invariant and gauge invariant, and has the correct behavior under soft limits. Moreover ${\cal P}_n$ has very interesting properties in four dimensions with a helicity configuration; it vanishes in exactly the only solution sector that ${\rm Pf}' \Psi_n$ is non-vanishing \eqref{F3R3}, and it is natural to divide the remaining sectors to obtain self-dual and anti-self-dual parts of $F^3$ and $R^3$ amplitudes, \eqref{SD}.

$F^3$ is not supersymmetrizable, which distinguishes it from any higher-dimensional operators in open superstring effective action. As shown in~\cite{Mafra:2011nv}, the polarization-dependence of open superstring amplitudes is encoded in $(n{-}3)!$ Yang-Mills color-ordered amplitudes:
\be
{\cal M}_n^{\rm open}(1,2,\ldots,n; \alpha')=\sum_{\rho \in S_{n{-}3}}~F^\rho_{1,2,\ldots,n}[\alpha']~M^{\rm YM}_n (\rho)\,,
\ee
where the sum is over $(n-3)!$ orderings, with scalar coefficients $F$'s containing the full $\alpha'$-dependence. Therefore, at any order in the $\alpha'$-expansion (see~\cite{Broedel:2013aza}), the amplitude always admits a CHY representation with the reduced Pfaffian ${\rm Pf}' \Psi_n$ (times a linear combination of Parke-Taylor factors). To give a very nice example, let's work out the CHY integrand for gluon amplitude from $F^4$ operator at ${\cal O}(\alpha'^2)$ of the open superstring effective action.

This is the first supersymmetrizable correction to Yang-Mills theory, and the amplitude with one insertion of $F^4$ has been studied in four dimensions~\cite{Stieberger:2006te} and in general dimensions~\cite{Mafra:2012kh} from superstring theory. There is also an interesting observation that MHV $F^4$ amplitude is proportional to the famous all-plus amplitude at one-loop level~\cite{BjerrumBohr:2011xe}. It turns out that $F^4$ color-ordered amplitude have a remarkably compact CHY formula
\be
{\cal I}_n^{F^4_{\rm s.s.}} (1,2,\ldots,n)=\left(\sum_{i<j<k<l} \sigma_{i j} \frac{s_{j,k}}{\sigma_{j k}} \sigma_{k l} \frac{s_{l i}}{\sigma_{l i}}\right)\,{\rm PT}(1,2,\ldots,n)~{\rm Pf}'\Psi_n\,.
\ee
which has been verified against the all-multiplicity result in~\cite{Mafra:2014oia}. The prefactor can be viewed as a CHY $D$-dimension ``uplift" of the spinor numerator $\sum \langle i\,j\rangle [j\,k] \langle k\,l\rangle [l\,i]$ for all-plus/MHV $F^4$ amplitude, which the formula reduces to for MHV helicities.

To all orders in $\alpha'$, no new ingredient for polarizations is needed for amplitudes from any operator from superstrings, which is in sharp contrast with ${\cal P}_n$ for $F^3$ amplitude! Our results for $F^3$ amplitudes and the double copies may open up a new direction for encoding more higher-dimensional operators in CHY formulation. Among other things, this can shed new lights in understanding amplitudes from the bosonic string effective action along the line of~\cite{Huang:2016tag}. Besides, from our formulas we can obtain BCJ numerators for $F^3$ amplitudes, similar to the Yang-Mills case in \cite{Cachazo:2013iea}. Along this line (also see \cite{Mafra:2011kj} from string theory), we hope to understand better the color/kinematics duality for $F^3$ and beyond.

The most intriguing feature of the new object ${\cal P}_n$ is its properties when reduced to four dimensions. With the only exception of bi-adjoint scalar theory~\cite{Cachazo:2016sdc}, every CHY formula so far is only non-vanishing in one sector of the 4d scattering equations (for given helicities), and they all nicely correspond to ambitwistor string theory~\cite{Mason:2013sva} with worldsheet supersymmetries~\cite{Ohmori:2015sha, Casali:2015vta}. ${\cal P}_n$ is totally different and it is likely to correspond to correlators from some bosonic version of the worldsheet models. It would be highly desirable to find such models. It would also be very interesting to see how these features in four dimensions can be derived from some four-dimensional ambitwistor string models directly~\cite{Geyer:2014fka}.

Our formula in gauge theory is for gluon amplitudes with a single insertion of $F^3$ operator, so it is also a formula for form factors in the soft limit. In the limit, it can be viewed as a very non-trivial generalization of earlier four-dimensional results on form factors for $F^2$ operator~\cite{He:2016jdg} and those in ${\cal N}=4$ SYM~\cite{He:2016dol}. An outstanding open question in this direction is about extending the construction to include multiple insertions of operators. Last but not least, recently there has been progress on loop integrands from scattering equations~\cite{Adamo:2013tsa,Geyer:2015jch,Geyer:2015bja,He:2015yua,Cachazo:2015aol}, and it would be highly desirable to see if our results can shed new lights on obtaining {\it integrated} loop amplitudes in this formulation. In particular, the amplitudes we studied here can be considered as counterterms for UV divergences of such loop amplitudes (see very interesting recent studies of Gauss-Bonnet term in quantum gravity~\cite{Bern:2015xsa, Cheung:2016wjt}), and it certainly deserves further investigations along this direction.

\section*{Acknowledgments}
S.H. thanks Nima Arkani-Hamed, Freddy Cachazo, Yu-tin Huang and Oliver Schlotterer for useful discussions. S.H.'s research is supported in part by the Thousand Young Talents program and the Key Research Program of Frontier Sciences of CAS (Grant No. QYZDBSSW-SYS014). Y.Z.'s research is partly supported by NSFC Grants No. 11375026.

\appendix
\section{Details for Reducing the Formulas to Four Dimensions}
As discussed in  ~\cite{He:2016vfi}, the scattering equations, \eqref{scatt}, were originally derived as the null condition $p^2(z)=0$ with $  p^\mu(z) \,:=\, \sum_{a=1}^n\, k_a^\mu~\prod_{b\neq a}\,(z-\sigma_b)$. In four dimensions, it is equivalent to the existence of polynomials
$\lambda(z)$ and $\tilde\lambda(z)$ with their degree added up to $(n{-}2)$, such that $p^{\alpha\dot\alpha}(z)=\lambda^\alpha(z) \tilde\lambda^{\dot\alpha}(z)$. Then we get $n-3$ different sets of 4d equations, with the degree of $\lambda(z)$ donated as $k'{-}1$ equal to $1,\cdots,n{-}3$ separately,
 \begin{align}\label{4dr}
\tilde\lambda_b^{\dot{\alpha}}  - \sum_{p\in +'} {\tilde\lambda^{\dot{\alpha}}_p \over (b\,p)} \,=\, 0
  ~~\text{for}~b\in -',
  \quad
 \lambda_p^\alpha - \sum_{b\in -'} {\lambda_b^\alpha \over (p\,b)} \,=\, 0
  ~~\text{for}~p=+',
\end{align}
Here $-'$ and $+'$ are arbitrary two sets of the $n$ external particles, with their length equal to $k'$ and $n-k'$ separately.  The variables are $\sigma$'s and $t$'s, which can be combined into $n$ variables in $\mathbb{C}^2$, $\sigma^{\alpha}_a=\frac 1{t_a} (\sigma_a, 1)$, and the two bracket is defined as $\displaystyle (a\,b) := (\sigma_a-\sigma_b)/(t_at_b)$.

Each solution of \eqref{scatt} corresponds to a unique solution $\{\sigma_a, t_a\}$ of \eqref{4dr} for some $k'$, with identical cross-ratios of the $\sigma$'s. For each $k'$, \eqref{4dr} have Eulerian number of solutions, $E_{n{-}3,k'{-}2}$, and the union of them for all sectors give $(n{-}3)!$ solutions of \eqref{scatt}, with $(n{-}3)!=\sum_{k'=2}^{n{-}2} E_{n{-}3,k'{-}2}$~\cite{Cachazo:2013iaa}. When reducing CHY formulas to 4d, it is convenient to view \eqref{4dr} as a change of variables: we refer to $\lambda_{I=1,\ldots,k}$, $\tilde\lambda_{i=k{+}1,\ldots,n}$ and $t_a, \sigma_a$ as ``data" and \eqref{4dr} as writing $\lambda_{i=k{+}1,\ldots, n}$ and $\tilde\lambda_{I=1,\ldots, k}$ in terms of the data. This is equivalent to evaluation on the support of solutions in the $k'$ sector.

Based on these considerations, now we derive the explicit expression when reducing $C_{aa}$ to four dimensions.  When $a\in -$ and  $a\in -'$,
\ba
C_{aa}=-\sum_{{b\in {\rm -'},\,b\neq a}}\frac{\<ab\>[b\mu]}{[a\mu]\s_{ab}}
-\sum_{p\in {\rm +'}}\frac{\<ap\>[p\mu]}{[a\mu]\s_{ap}}
\ea
By plugging in the solutions in $k'$ sector, or equivalently a change of variable, we have
\ba\label{a21}
C_{aa}&=&-\frac{1}{[a\mu]}
\sum_{b\neq a;\,p}
\Big(\frac{\<a b\>t_bt_p[p\mu]}{\s_{bp}\s_{ab}}+\frac{\<a b\>t_bt_p[p\mu]}{\s_{pb}\s_{ap}}
\Big)\nl
&=&-\frac{1}{[a\mu]}
\sum_{b\neq a;\,p}
\frac{\<a b\>t_bt_p[p\mu]}{\s_{ab}\s_{ap}}
\ea
In the second equality, we have taken the denominators together. Now it factorizes to two factors $-\sum_{b\neq a}
\frac{t_b\<a b\>}{t_a\s_{ab}}$ and $\sum_{p}\frac{t_at_p[p\mu]}{\s_{ap}[a\mu]}$, where the latter equals to $1$ according to 4d scattering equations~\eqref{4dr}. Finally,
\ba\label{a22}
C_{aa}=-\sum_{{b\in {\rm -'},\,b\neq a}}
\frac{t_b\<a b\>}{t_a\s_{ab}}
\ea
While $a\in -$ but  $a\notin -'$,
\ba
C_{aa}=&&-\sum_{{p\in {\rm +'},\,p\neq a}}\frac{\<ap\>[p\mu]}{[a\mu]\s_{ap}}
-\sum_{b\in {\rm -'}}\frac{\<ab\>[b\mu]}{[a\mu]\s_{ab}}
\ea
Plug in changes of variables, and now it comes out an extra term as $p=a$ also contributes.
\ba
C_{aa}=&&-\frac{1}{[a\mu]}
\sum_{\substack{p\in {\rm +'},\,p\neq a\\b\in {\rm -'}}}
\Big(\frac{\<a b\>t_bt_p[p\mu]}{\s_{pb}\s_{ap}}+\frac{\<a b\>t_bt_p[p\mu]}{\s_{bp}\s_{ab}}
\Big)+\sum_{b\in {\rm -'}}\frac{\<ab\>t_bt_a}{\s_{ba}^2}
\ea
The first term on the RHS vanishes following the trick as in \eqref{a21}, \eqref{a22}, it becomes
\ba
-\frac{1}{[a\mu]}
\sum_{\substack{p\in {\rm +'},\,p\neq a\\b\in {\rm -'}}}
\frac{\<a b\>t_bt_p[p\mu]}{\s_{ab}\s_{ap}}=-\sum_{{b\in {\rm -'}}}
\frac{t_at_b\<a b\>}{\s_{ab}}
\sum_{p\in {\rm +'},\,p\neq a}\frac{t_p[p\mu]}{t_a\s_{ap}[a\mu]}=0\,,
\ea
then we see that $C_{aa}$ only has contribution from the extra term, and we obtain
\ba
C_{aa}
=\sum_{\substack{b< c\\b,c\in {\rm -'}}}\frac{\<cb\>t_bt_ct_a^2\s_{bc}}{\s_{ba}^2\s_{ca}^2}\,.
\ea
Similarly we can work out the other two cases, and the final result is
\ba\label{Cii}
C_{aa}=
\begin{cases}
-\sum\limits_{b\neq a;\ b\in{\rm -'}}\frac{t_b}{t_a}\frac{\<ab\>}{\s_{ab}}\quad &a\in- \; \text{and}\; a\in {\rm -'}
\\
-t_a^2\sum\limits_{b<c;\ b,c\in {\rm -'}}\frac{t_b t_c \s_{bc}\<bc\>}{\s_{ab}^2\s_{ac}^2}\quad &a\in -\; \text{but}\; a\notin {\rm -'}\;
\\
-\sum\limits_{b\neq a;\ b\in{\rm +'}}\frac{t_b}{t_a}\frac{[ab]}{\s_{ab}}\quad &a\in+\; \text{and}\; a\in{\rm +'}
\\
-t_a^2\sum\limits_{b<c;\ b,c\in {\rm +'}}\frac{t_b t_c \s_{bc}[bc]}{\s_{ab}^2\s_{ac}^2}\quad &a\in +\; \text{but}\; a\notin {\rm +'}\;.
\end{cases}
\ea
Thus we can define the diagonal elements of the two (generalized) matrices ${\bm h}_{k}^{k'}$ and ${\bm {\tilde{h}}}_{n-k}^{k'}$ as
\ba\label{haakp}
{h}_{aa}&=&-\sum\limits_{b\neq a;\ b\in{\rm -'}}\frac{t_b}{t_a}\frac{\<ab\>}{\s_{ab}}\quad\qquad a\in- \; \text{and}\; a\in {\rm -'}\nl
{h}_{aa}&=&-t_a^2\sum\limits_{b<c;\ b,c\in {\rm -'}}\frac{t_b t_c \s_{bc}\<bc\>}{\s_{ab}^2\s_{ac}^2}\quad a\in -\; \text{but}\; a\notin {\rm -'}\;
\ea
and
\ba
\tilde{{h}}_{aa}&=&-\sum\limits_{b\neq a;\ b\in{\rm +'}}\frac{t_b}{t_a}\frac{[ab]}{\s_{ab}}\quad\qquad a\in+\; \text{and}\; a\in{\rm +'}\nl
\tilde{{h}}_{aa}&=&-t_a^2\sum\limits_{b<c;\ b,c\in {\rm +'}}\frac{t_b t_c \s_{bc}[bc]}{\s_{ab}^2\s_{ac}^2}\quad a\in +\; \text{but}\; a\notin {\rm +'}\;.
\ea
Considering the special case of ${\cal M}_{n,3}^{F_+^3}$, we find that ${\cal P}_n$ reduces to
\ba
{\cal P}_n=h_{pp}h_{qq}t_r^2\frac{t_p t_q \s_{pq}\<pq\>}{\s_{rp}^2\s_{rq}^2}\mathrm{det}\,{\bm {\tilde{h}}}_{n-3}^{2}\; \qquad {\rm when}\ k=3,k'=2\;.
\ea
where the diagonal elements of the matrix ${\bm {\tilde{h}}}_{n-3}^{2}$ are given by
\ba
\tilde{{h}}_{aa}&=&-\sum\limits_{b\neq a,p,q}\frac{t_b}{t_a}\frac{[ab]}{\s_{ab}}\,,
\ea
Thus we have seen that ${\bm {\tilde{h}}}_{n-3}^{2}$ is nothing but the reduced matrix $|{\bm{\tilde{h}}}_{n-2}|_r^r$.

The 4d formula for the self-dual $F^3$ amplitude with $k=3$ now reads
\ba
M_{n,3}^{{ F}^3_+}=
\<pq\>^2
\int\frac{d^{2n}\s}{{\rm vol~GL}(2,\mathbb{C})}
\prod_{a\neq p,q}\d^2
\left(
{\l}_a-\frac{{\l}_p}{(ap)}-\frac{{\l}_q}{(aq)}
\right)\nl
\times\frac{1}{(12)(23)\cdots(n1)}
\frac{h_{pp}h_{qq}t_r^2\frac{t_p t_q \s_{pq}\<pq\>}{\s_{rp}^2\s_{rq}^2}\mathrm{det}\,{\bm {\tilde{h}}}_{n-3}^{2}}
{\mathrm{det'}\,{\bm{h}}_{2}\mathrm{det'}\,{\bm{\tilde{h}}}_{n-2}}
\ea
Here the four delta functions for $p,q$ in \eqref{4dr} has been stripped as a delta function of momentum conversation $\d^4{(P)}$ and dropped, which leads to the prefactor $\<pq\>^2$. The two reduced determinants $\mathrm{det'}\,{\bm{h}}_{2}$, $\mathrm{det'}\,{\bm{\tilde{h}}}_{n-2}$ come as the Jacobian from rewriting scattering equations \eqref{scatt} to 4d scattering equations \eqref{4dr}. Luckily, $\mathrm{det'}\,{\bm{\tilde{h}}}_{n-2}$ cancels $\mathrm{det}\,{\bm {\tilde{h}}}_{n-3}^{2}$ in the numerator since the two matrices ${\bm {\tilde{h}}}_{n-3}^{2}$  and $|{\bm{\tilde{h}}}_{n-2}|_r^r$ are identical:
\ba
M_{n,3}^{{ F}^3_+}&=&\frac{\<pq\>^2}{J_{n,2}}
\frac{1}{(12)(23)\cdots(n1)}
\frac{h_{pp}h_{qq}t_r^2\frac{t_p t_q \s_{pq}\<pq\>}{\s_{rp}^2\s_{rq}^2}}
{\mathrm{det'}\,{\bm{h}}_{2}\frac{1}{t_r^2}}\nl
&=&
\frac{\<pq\>^2}{J_{n,2}}
\frac{1}{(12)(23)\cdots(n1)}\frac{\<pq\>^2}
{(rp)^2(rq)^2}
\ea
Here $J_{n,2}$ is the Jacobian of 4d scattering equations.

The unique solutions for the scattering equations of sector $k'=2$ is given by:
\ba\label{MHVsol}
(ap)=\frac{\<pq\>}{\<aq\>},\quad
(aq)=\frac{\<pq\>}{\<pa\>},\quad
(ab)=\frac{\<pq\>^3\<ab\>}{\<pa\>\<qa\>\<pb\>\<qb\>},\quad
(pq)=1.
\ea
Evaluating on this solution we obtain
\ba
\frac{1}{(12)(23)\cdots(n1)}
=\frac{\prod_{a\neq p,q}\<pa\>^2\<qa\>^2}{\<pq\>^{3n-8}\<12\>\<23\>\cdots\<n1\>}
\qquad
\frac{\<pq\>^2}
{(rp)^2(rq)^2}
=\frac{\<rp\>^2\<rq\>^2}{\<pq\>^2}
\ea
and the Jacobian of of 4d scattering equations becomes
\ba
J_{n,2}=\frac{\prod_{a\neq p,q}\<pa\>^2\<qa\>^2}{\<pq\>^{3n-6}}\,.
\ea
Combining these factors we obtain the ``Parke-Taylor-like" formula in the main text.
\bibliographystyle{unsrt}
\bibliography{mybibliography}

\end{document}